\documentclass[10pt,a4paper,oneside]{article}

\usepackage[utf8]{inputenc} 
\usepackage[T1]{fontenc}
\usepackage{lmodern}
\usepackage{url}            
\usepackage{booktabs}       
\usepackage{amsfonts}       
\usepackage{nicefrac}       
\usepackage[a4paper,left=3cm,right=2cm,top=2.5cm,bottom=2.5cm]{geometry}
\usepackage{graphicx}
\usepackage{tabularx}
\usepackage{amsmath}
\usepackage{braket}
\usepackage{amsthm}
\usepackage{amsbsy}
\usepackage{amssymb}
\usepackage{doi}
\title{Epitaxial thin films of binary Eu-compounds close to a valence transition}

\author{Sebastian K\"olsch\,\footnote{Corresponding author. E-mail: \href{mailto:koelsch@physik.uni-frankfurt.de}{koelsch@physik.uni-frankfurt.de}} , Alfons Schuck, Michael Huth}
\date{%
	Thin films and nanostructures, Physical Institute, Goethe University Frankfurt,
	Max-von-Laue Street 1, Frankfurt am Main 60438, Germany\\%
}
\begin{document}
	\maketitle
	\vspace{-6mm}
	\begin{abstract}
		Intermetallic binary compounds of europium reveal a variety of interesting phenomena due to the interconnection between two different magnetic and 4f electronic (valence) states, which are particularly close in energy.
		The valence states or magnetic properties are thus particularly sensitive to strain-tuning in these materials. 
		Consequently, we grew epitaxial EuPd$_2$ (magnetic Eu$^{2+}$) and EuPd$_3$ (nonmagnetic Eu$^{3+}$) thin films on MgO(001) substrates using molecular beam epitaxy.
		Ambient X-ray diffraction confirms an epitaxial relationship of cubic Laves-type (C15) EuPd$_2$ with an (111)-out-of-plane orientation, whereby eight distinct in-plane crystallographic domains develop.
		For simple cubic EuPd$_3$ two different out-of-plane orientations can be obtained by changing the substrate annealing temperature under ultra-high vacuum conditions from 600\,°C to 1000\,°C for one hour.
		A small resistance minimum evolves for EuPd$_3$ thin films grown with low temperature substrate annealing, which was previously found even in single crystals of EuPd$_3$ and might be attributed to a Kondo or weak localization effect.
		Absence of influence of an applied magnetic fields and magnetotransport measurements suggest a nonmagnetic ground state for EuPd$_3$ thin films, i.\,e., a purely trivalent Eu valence, as found in EuPd$_3$ single crystals.
		For EuPd$_2$ magnetic ordering below $\sim$72\,K is observed, quite similar to single crystal behavior.
		Field dependent measurements of the magnetoresistance and the Hall effect show hysteresis effects below $\sim$0.4\,T and an anomalous Hall effect below $\sim$70\,K, which saturates around 1.4\,T, thus proving a ferromagnetic ground state of the divalent Eu.
	\end{abstract}
	
	\section{Introduction}
	Intermetallic binary and ternary compounds of europium attracted much interest in recent years, due to a rich variety of characteristic features, attributed to the 4f-electrons \cite{onuki_divalent_2017}.
	In the case of Eu$^{2+}$ (4f$^7$ configuration) the spin (S = 7/2), orbital (L = 0) and total (J = 7/2) angular momenta can be derived according to Hund's rule, yielding a high magnetic moment of 7.94\,$\mu_B$/Eu$^{2+}$, attributed only to the spins of the localized 4f-electrons \cite{wickman_moessbauer_1967}.
	In contrast, trivalent Eu$^{3+}$ (4f$^6$) has a vanishing magnetic moment (S = L = 3, J = 0) in the ground state.
	Furthermore, depending on the valence state strongly different ionic radii occur, whereby the Eu$^{2+}$ ion is about 20\% bigger than Eu$^{3+}$, depending on the coordination number \cite{shannon_revised_1976}.
	The Eu valence in intermetallic compounds thus depends on the local environment of the Eu-ions, as well as on the number and kind of nearest neighbors.
	As discussed by Doniach \cite{doniach_phase_1977}, the competition between the Ruderman-Kittel-Kasuya-Yosida (RKKY) interaction and the Kondo-effect suggests a generic p-T-phase diagram, which is greatly influenced by the tunable magnetic exchange interaction parameter J$_{cf}$ between the conduction and 4f electrons.
	Interestingly, for Eu both valence states can be rather close in energy, leading to a variety of competing phenomena such as ferro-/antiferromagnetic ordering, Kondo-effect or valence crossover \cite{onuki_divalent_2017}. 
	In the prototypical ternary compound EuPd$_2$Si$_2$ the valence-fluctuation is known to be accompanied by a large and abrupt reduction ($\sim$2\%) of the tetragonal a-lattice parameter \cite{sampathkumaran_new_1981}, implying a substantial interconnection between lattice and electronic degrees of freedom \cite{kliemt_strong_2022}.
	In particular, a valence transition from Eu$^{2+}\rightarrow$ Eu$^{3+}$ may be tuned by temperature \cite{sampathkumaran_new_1981}, high magnetic fields \cite{wada_first-order_1996}, hydrostatic \cite{adams_effect_1991} or chemical pressure due to, e.g., isoelectronic substitution of palladium by platinum in Eu(Pd$_{2-x}$Pt$_{x}$)Si$_2$ \cite{mitsuda_field-induced_1997}.
	Recently, we have shown that it is possible to suppress the valence transition in epitaxial and highly crystalline EuPd$_2$Si$_2$(001)/MgO(001) thin films due to the particularly strong coupling between the electronic fluctuations and the lattice degrees of freedom via a mechanical clamping effect to the stiff MgO substrate \cite{koelsch_clamping_2022}.
	Poly- and single crystals of both binary compounds, EuPd$_2$ and EuPd$_3$, show stable divalent and trivalent europium valencies down to lowest temperatures, respectively, as explored, e.g., by Mössbauer \cite{wickman_moessbauer_1967, harris_study_1971} or nuclear magnetic resonance (NMR) \cite{kropp_electric_1979} spectroscopic studies.
	In the case of EuPd$_2$, ferromagnetic ordering of Eu is observed below $\sim$80\,K \cite{harris_study_1971} and the Curie temperature can be raised linearly by the application of hydrostatic pressure \cite{nakamura_magnetic_2016}.
	However, even in pressures up to 8\,GPa the europium remains divalent in EuPd$_2$. 
	Previous NMR studies on EuPd$_2$ polycrystals revealed, that magnetic ordering is promoted by indirect RKKY exchange interaction through polarization of the 6s$^2$ conduction electrons, donated from Eu \cite{kropp_indirect_1979}. 
	
	In contradistinction, poly-\cite{harris_study_1971} and single\cite{takeuchi_contribution_2014} crystal studies of EuPd$_3$ confirm a nonmagnetic ground state.
	Due to the thermal occupation of close-lying excited states of Eu$^{3+}$ a weak and nearly temperature-independent Van Vleck susceptibility is experimentally observed in EuPd$_3$ with the trivalent Eu-state \cite{gardner_the_1972}.
	As possible reason for a small upturn of resistivity upon cooling below 10\,K in EuPd$_3$ single crystals, exhibiting a slight excess of $\sim$0.1\% divalent Eu, Kondo-like behavior due to magnetic impurities was suggested \cite{takeuchi_contribution_2014}.
	As evident from previous photoemission studies of annealed europium films on palladium single crystals, most europium-palladium compounds contain some divalent Eu in the outermost layers independent of temperature \cite{bertran_growth_1991, bertran_growth_1992}, indicating different properties at the surface as compared to the bulk \cite{wertheim_final-state_1985, bertran_the_1992}.
	Unfortunately, strong interdiffusion effects even at liquid nitrogen temperatures inhibit the investigation of stoichiometric and epitaxial europium-palladium films if a palladium substrate is chosen \cite{bertran_growth_1992}. 
	Concerning epitaxial europium-palladium thin films we know of only one report, in which the codeposition of ultra thin films of EuPd$_3$ (<4\,nm) is investigated on MgO(001) substrates and Fe(001) templates by electron diffraction, photoelectron spectroscopy and ferromagnetic resonance experiments \cite{maslankiewicz_europium-palladium_2008}.
	The authors conclude the growth mode on 600\,°C annealed MgO(001) to be of the Vollmer-Weber type based on electron diffraction in-plane studies, where at first random in-plane (111)-textured islands coalesce and finally turn into a EuPd$_3$(001)-film.
	This growth mode is supported by the similarity between fcc Pd and sc EuPd$_3$ (replacing palladium atoms in the corners of the fcc Pd unit cell by europium atoms yields the AuCu$_3$ structure of EuPd$_3$), whereby Pd grows island-like on MgO(001) \cite{renaud_growth_1999}.
	As evidenced from angle-resolved X-ray photoelectron spectroscopy experiments only a small fraction of Eu is divalent at room temperature \cite{maslankiewicz_europium-palladium_2008}.
	To our knowledge, up to now no epitaxial thin film specific report regarding EuPd$_2$ on non-palladium substrates has been published, which leaves several questions unanswered.
	Furthermore, due to experimental difficulties for many RE-based compounds synthesis of phase-pure epitaxial thin films has not yet been achieved successfully \cite{chatterjee_heavy_2021}.
	In this study we report experimental results concerning the growth of epitaxial EuPd$_3$ and EuPd$_2$ thin films (thickness $\sim$ 50\,nm), as well as their structural and temperature-dependent magnetotransport properties.
	
	\section{Experimental details}
	Epitaxially grown thin films were prepared on MgO(001) substrates using molecular beam epitaxy (MBE) in an ultra-high vacuum chamber described elsewhere \cite{koelsch_clamping_2022}.
	During growth, the absolute pressure in the chamber did not exceed $5\times 10^{-9}$\,mbar and was composed mainly ($\sim$90\%) of hydrogen, due to the gas release from the heated europium source.
	Residual gas analysis was performed with a quadrupole mass spectrometer from Pfeiffer Balzers (model QMA120A).
	According to the results from our former epitaxial growth study on MgO substrates \cite{koelsch_clamping_2022}, no chemical cleaning of the substrate was performed in order to reduce possible growth-disturbing hydroxylation \cite{febvrier_wet_2017} and subsequent attack of the surface \cite{duriez_structural_1990}.
	Instead, the substrates were degassed at 450\,°C for one hour and thermally cleaned at elevated temperature prior to growth as described below.
	For the deposition of Pd an electron beam evaporator was used, whose material flux was feedback-controlled by a water-cooled quartz crystal microbalance.
	Eu was sublimated from a pyrolytic boron nitride (PBN) crucible inside an effusioncell at a temperature between 440-470$^\circ$C, with a stability better than 0.1$^\circ$C.
	It is important to note that the substrate temperature during codeposition was held constant at 450\,°C and is comparable to the sublimation temperature of europium.
	Significant desorption may therefore take place at the surface and will influence the thin film growth, as discussed below.
	After codeposition the samples were allowed to cool to room temperature before an amorphous Si capping layer with a typical thickness of 5\,nm was deposited using another electron beam evaporator to prevent further oxidation due to the highly reactive nature of europium.
	Before and after growth the surface crystallinity was inspected via reflection high-energy electron diffraction (RHEED) with a 15\,keV electron source.
	No signs of crystallinity could be observed via RHEED after the deposition of the silicon film.
	Scanning electron microscopy (SEM) images were acquired ex-situ in a FEI Nova NanoLab 600 to study the surface morphology.
	Structural characterization via symmetric and asymmetric X-ray diffraction (XRD) was done with a Bruker D8 Discover high-resolution diffractometer using Cu$_{K, \alpha}$ radiation with a parallelized primary beam and a diffracted-side monochromator in air.
	For analysis of the X-ray scans, revealing high and low angle oscillations, Bruker$^\prime$s DiffracPlus Leptos Software was used.

	\section{Results and discussion}
	\subsection{EuPd$_3$/MgO(001)}
	
	\subsubsection{Growth mode and structural properties}
	
	\begin{figure}[htb]
		\includegraphics[width=\textwidth]{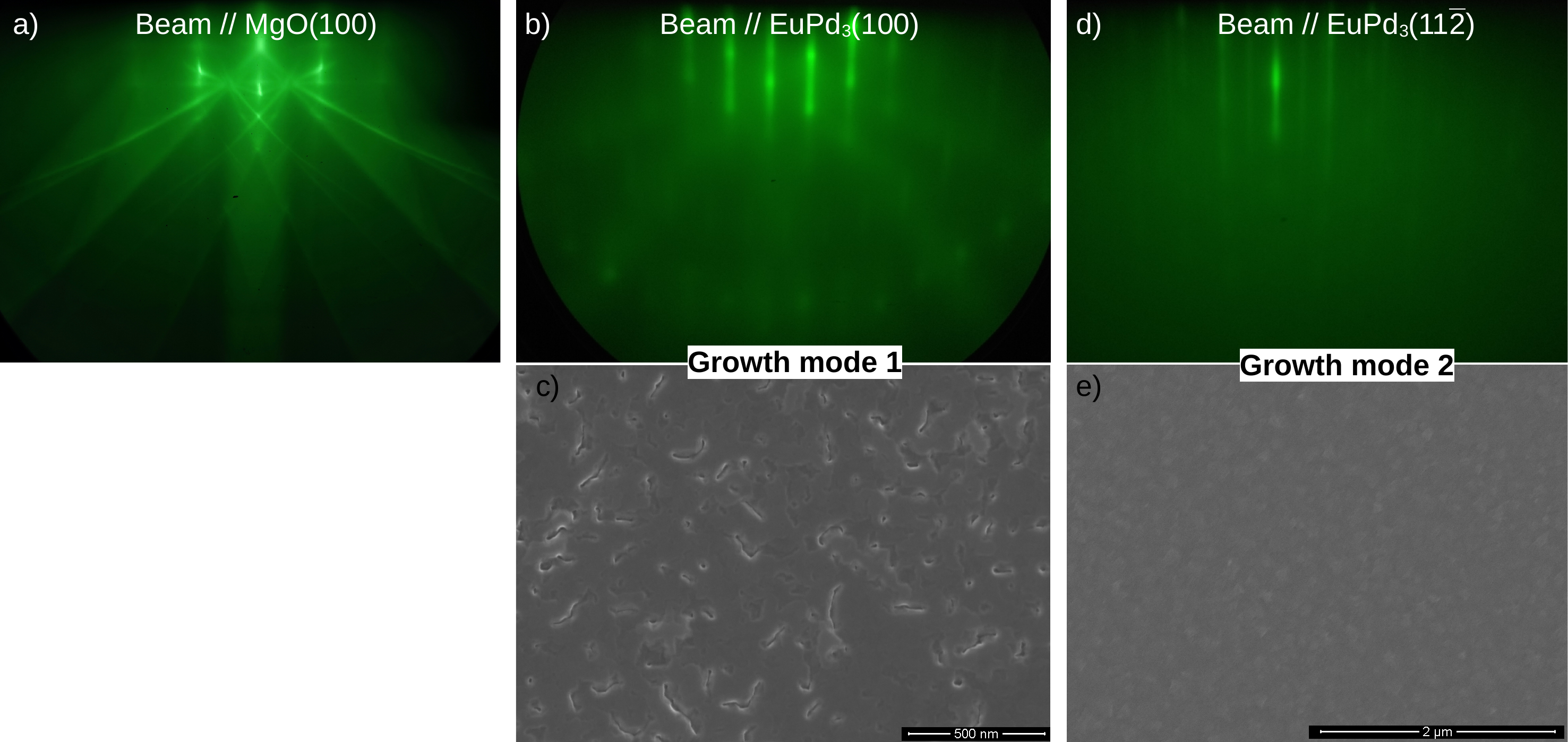}
		\caption{a) RHEED image of the pure MgO(001) substrate with beam along the [100] direction, directly before codeposition. b) and d) RHEED images after thin film deposition and before Si capping, grown in mode 1 and mode 2 without any change of sample position. c) and e) corresponding SEM images of the EuPd$_3$ films including the Si capping layer. The scale bar corresponds to 500\,nm and 2\,$\mu$m, respectively.}
		\label{rheed-eupd3}
	\end{figure}
	
	In comparison to the cubic lattice parameter of fcc MgO a = 4.212\,\AA\, \cite{smith_low-temperature_1968}, EuPd$_3$ has a simple cubic structure with a = 4.09\,\AA\ \cite{villars_pearsons_1991}. 
	Calculation of the lattice mismatch at room temperature thus yields about -3\%, i.\,e. a small tensile strain for an epitaxial EuPd$_3$ film may be expected.
	As reported before, low temperature annealing of the MgO substrate just before codeposition leads to the first growth mode of EuPd$_3$ on MgO(001) \cite{maslankiewicz_europium-palladium_2008}, here called mode 1.
	Upon rotation of the pure MgO substrate after annealing at 600$^\circ$C around the surface normal, RHEED azimuthal scans show a fourfold symmetry, see Fig.\,\ref{rheed-eupd3}\,a). 
	Subsequent to the codeposition of the EuPd$_3$ thin film the same fourfold symmetry is observed, see Fig.\,\ref{rheed-eupd3}\,b).
	Without any sample rotation the RHEED pattern of the thin film shows equidistant long streaks and reflexes also in the first order Laue circle, indicative of high structural in-plane order.
	Comparison of the directions for the main symmetry axes from the substrate and the thin film implies a parallel alignment of their crystallographic a-axes and a simple cube-on-cube growth model can be stated. 
	Furthermore, broad Kikuchi bands are clearly visible, pointing to a laterally well-ordered crystalline film on the length scale of several hundred nanometers.
	SEM (Fig.\,\ref{rheed-eupd3} c) images of the Si-capped thin films reveal a rather smooth and flat topography with some valleys and holes on a lateral scale of about 50-200\,nm.\\
	
	Choosing a substrate annealing temperature of 1000\,°C for one hour leads to another growth mode, herein called mode 2.
	Such prepared MgO surfaces show nearly the same RHEED patterns as before.
	Direct observation of the RHEED pattern after thin film growth reveals however less contrast and only very long streaks near the zeroth order Laue circle.
	In addition, without any sample rotation weak lines between stronger main lines appear, see Fig.\,\ref{rheed-eupd3} d).
	Furthermore, no Kikuchi bands or lines develop, implying less in-plane order in growth mode 2.
	Interestingly, rotation of the sample yields the same pattern every 60°.
	SEM images (see Fig.\,\ref{rheed-eupd3} e) only exhibit weak contrast, without any visible holes or valleys.
	Instead, some small triangular facets can be observed on close inspection, indicating a very flat thin film with strongly coalesced (111)-oriented islands.\\
	
	\begin{figure*}[htb]
		\includegraphics[width=\textwidth]{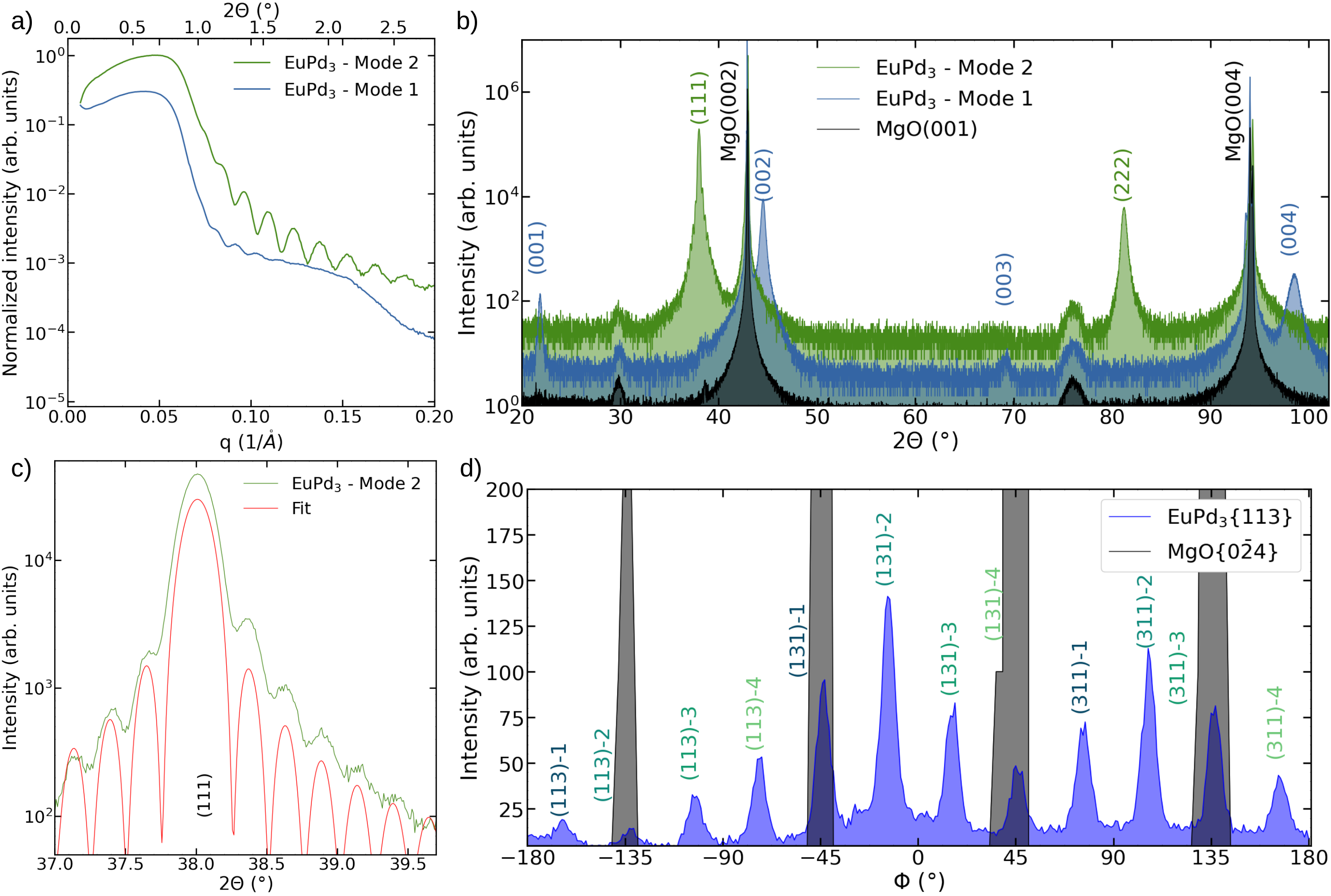}
		\caption{a) X-ray reflectometry scan of two EuPd$_3$ thin films, grown in mode 1 (blue) or 2 (green). Curves are offset for clarity. b) Wide angle symmetric scan of two different EuPd$_3$ thin films grown in mode 1 and 2, respectively. A scan of native MgO(001) substrate is shown for reference. c) Magnification of a longitudinal symmetric X-ray diffractogram near the EuPd$_3$(111)-reflex, grown in mode 2. Fit (red) of the Laue oscillations is included. d) Asymmetric XRD scan of the MgO\{0$\bar{2}$4\}- and EuPd$_3$\{113\}-reflex families. See text for details.}
		\label{xrd-eupd3}
	\end{figure*}
	
	For both growth modes Kiessig fringes arise from a relatively smooth interface and surface, which are visible in symmetric low angle X-ray reflectometry (XRR) scans (Fig.\,\ref{xrd-eupd3} a) up to 2$\Theta$ = 3$^\circ$.
	From this an average film thickness d of 39\,$\pm\,1$\,nm for EuPd$_3$ in mode 1 and 50\,$\pm\,1$\,nm for mode 2 is deduced, respectively.
	For the second growth mode a slightly thicker film with longer codeposition time was chosen, which showed more oscillations, indicating less interfacial and surface roughness.
	The symmetric high angle X-ray diffraction scan (Fig.\,\ref{xrd-eupd3} c) reveals in mode 1 only (00$\ell$)-reflexes of EuPd$_3$ with natural number $\ell$ appearing up to the 4th order, suggesting a well-ordered epitaxial thin film. 
	In case of mode 2 only the EuPd$_3$(111)- and EuPd$_3$(222)-reflex occur with high intensity.
	Furthermore, Laue oscillations next to the EuPd$_3$(111)-reflex (Fig.\,\ref{xrd-eupd3} b) indicate a high degree of structural order for the out-of-plane direction with a crystalline coherence length L$_c\approx$ 37\,nm.
	This suggests $\sim$95\% crystalline volume fraction in the out-of-plane direction with respect to the total layer thickness of d = 39\,nm, which was obtained by analysis of the Kiessig fringes. 
	Around 76° a broad and small reflex is visible for all films and the pure MgO substrate, which appears due to scattering from the sample holder. 
	In both modes no crystalline impurity phase is detected, implying phase-pure epitaxial growth of either EuPd$_3$(100) or EuPd$_3$(111) on MgO(001), depending solely on the annealing temperature of the substrate and therefore the surface conditions prior to growth. 
	Additional asymmetric X-ray diffractograms (not shown) confirm the epitaxial relationship for the first growth mode as deduced from RHEED to be:
	
	\begin{center}
		(1): MgO\{100\}$\parallel$EuPd$_3$\{100\} \& MgO<001>$\parallel$EuPd$_3$<001>
	\end{center}
	
	In the high temperature annealing case (growth mode 2) instead an asymmetric XRD $\phi$-Scan (Fig.\,\ref{xrd-eupd3} d) of the EuPd$_3$\{113\}-reflex family reveals four equivalent rotational in-plane domains (labelled by different colors and numbers 1-4), which form the epitaxial relationship according to:
	\begin{center}
		(2): MgO\{100\}$\parallel$EuPd$_3$\{111\} \& MgO<010>$\parallel$EuPd$_3$<11$\bar{2}$>
	\end{center}

	\subsubsection{Magnetotransport properties}
	\begin{figure*}[htb]
		\includegraphics[width=\textwidth]{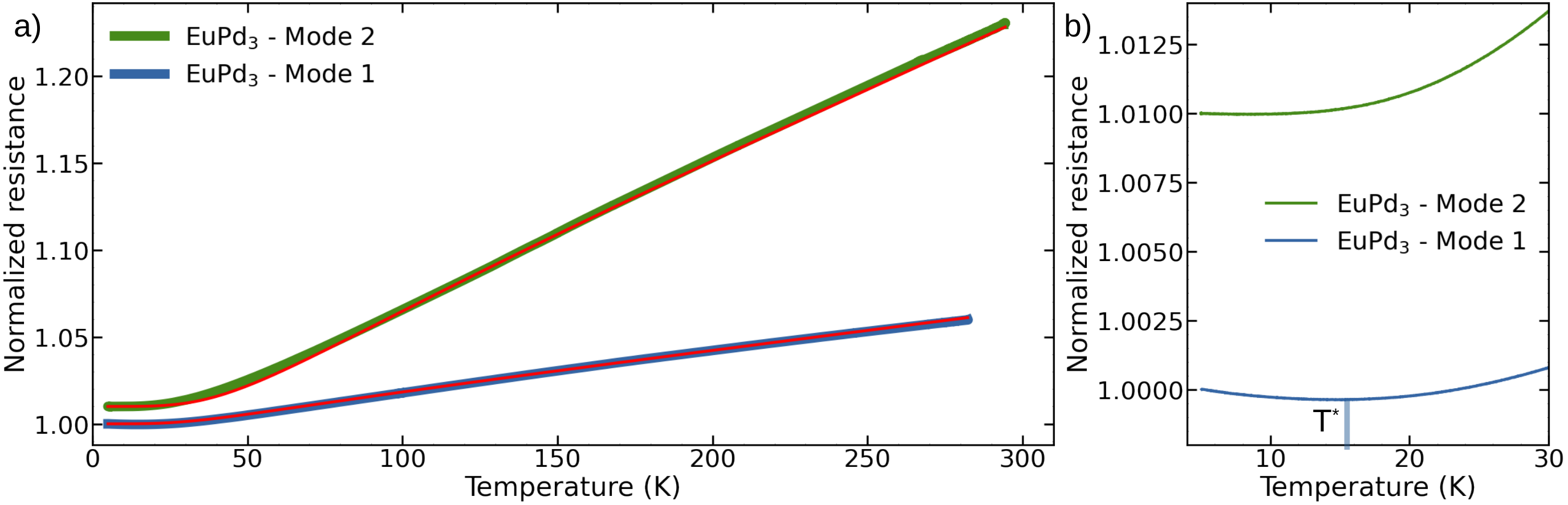}
		\caption{a) Resistance of the same two EuPd$_3$ films as before, normalized to the value at 5\,K. The upper curve is shifted for clarity. Included are fits (red) according to the Bloch-Gr\"uneisen model. b) Detailed view of the low temperature region, revealing a resistance minimum at T$^*$ around 15\,K for the first growth mode. See text for details.}
		\label{cryo-eupd3}
	\end{figure*}
	
	Patterning of thin films utilizing UV-photolithograpy and low-energy (<1000\,eV) Ar-ion sputtering were done to fabricate 6-contact Hall bar structures with a cross-area of 30\,$\times$\,100\,$\mu \text{m}^2$. 
	Low temperature magnetotransport measurements were performed inside a helium flow cryostat via a variable temperature insert between 3\,K and 300\,K in a magnetic field up to 5\,T.
	The measurements werde done at a constant current of 100\,$\mu$A, leading to current densities smaller than 1$\cdot10^8$\,A/m$^2$.
	For both growth modes, a simple metallic behavior below room temperature is visible in the normalized resistance, see Fig.\,\ref{cryo-eupd3} a).
	Interestingly, the calculated resistivity and residual resistivity ratios (RRR) values vary significantly, despite the quite similar thicknesses.
	For the film grown in mode 1 $\rho_1(T = 4\,$K) = 130\,$\mu\Omega$cm and RRR$_1\sim$ 1.06, whereas for growth mode 2 a much smaller resistivity of $\rho_2(4\,$K) = 37\,$\mu\Omega$cm and RRR$_2$ = 1.22 is calculated.
	Both properties thus reflect the higher structural order or crystallinity, which may be obtained by high temperature annealing of the substrate.
	Roughly above 50\,K the resistance scales for both variants linearly with respect to temperature and becomes nearly constant below about 20\,K.
	Furthermore, a resistivity minimum develops for the film in the first growth mode at a temperature of $T^*\approx 15\,$K, see Fig.\,\ref{cryo-eupd3} b).
	A similar resistance minimum is observed in well-ordered single crystals, with $\rho_0= 2.4\,\mu\Omega$cm and RRR = 3.1, which also show de Haas-van Alphen oscillations \cite{takeuchi_contribution_2014}.
	As there seems to be no contribution of higher J-multiplets of EuPd$_3$ to the resistivity, a simple Bloch-Gr\"uneisen model due to phonon scattering at elevated temperatures seems appropriate.
	In the bulk case this leads to a Debye temperature of $\Theta_D\sim$230\,K \cite{takeuchi_contribution_2014}.
	Evaluating this simple model to films of both growth modes shows good agreement, as can be seen from the concordance with the fits (red lines) in Fig.\,\ref{cryo-eupd3}\,a).
	Here Debye temperatures of 189\,K (mode 1) and 256\,K (mode 2) are obtained, which are as well comparable with $\Theta_D\sim$280\,K of pure palladium \cite{matula_electrical_1979}.
	In the case of lower crystallinity films in mode 1, the deviation from single crystal behavior is more pronounced, which can be seen, e.g., in $\rho_0$, RRR and $\Theta_D$. 
	Complementary measurements of the Hall effect (HE), i.\,e., the Hall resistance, always reveal a negative linear Hall coefficient with applied magnetic field in the out-of-plane direction for temperatures below 300\,K (not shown).
	Assuming a single-band model for the charge carriers thus leads to a nearly constant, electron-like charge carrier density of $\sim 6\cdot 10^{28}\,$m$^{-3}$.
	Furthermore, measurements of the longitudinal magnetoresistance (MR = $\left[\rho(B) - \rho(B = 0)\right]/\rho(B = 0)$) at low temperatures (< 10\,K) show a negligible but positive effect, i.\,e., a marginally larger resistivity of well below +0.1\% at 5\,T, as compared to the zero field value for films of both growth modes (not shown).
	
	\subsection{EuPd$_2$/MgO(001)}
	
	\subsubsection{Growth and structural properties}
	\begin{figure}[htb]
		\includegraphics[width=\textwidth]{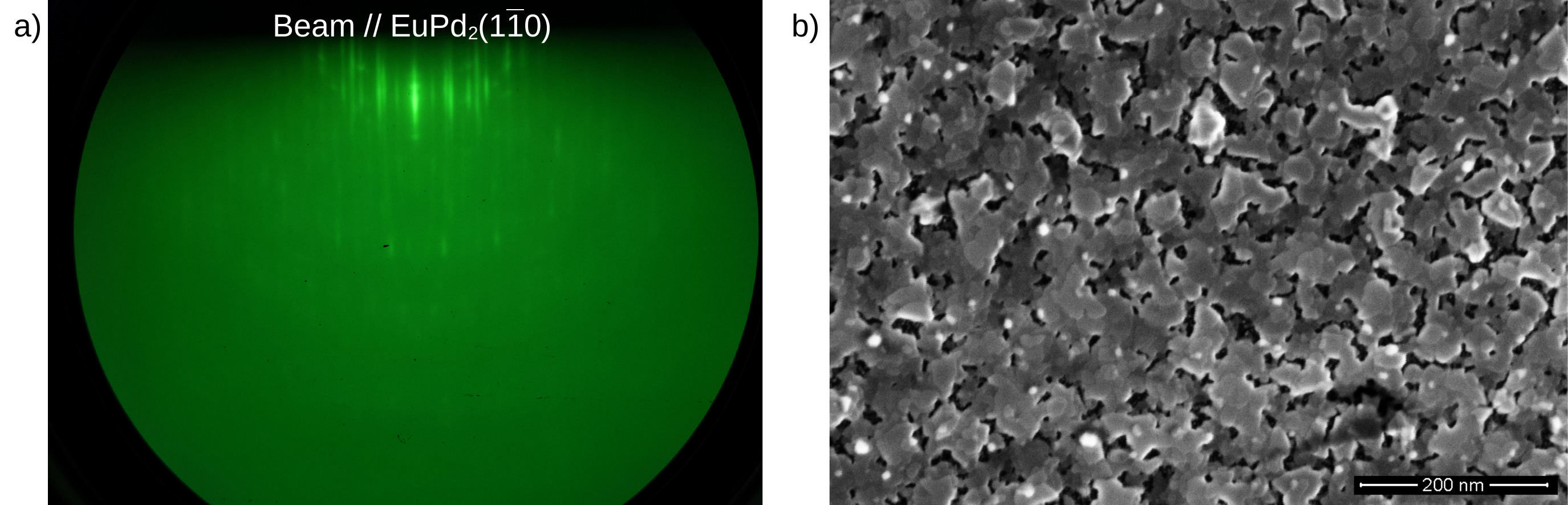}
		\caption{a) RHEED image subsequent to thin film deposition after rotation of the sample to a main symmetry direction of the pattern. b) Corresponding SEM image of the EuPd$_2$ film including the Si capping layer. The scale bar corresponds to 200\,nm.}
		\label{rheed-eupd2}
	\end{figure}
	
	Since EuPd$_2$ crystallizes in the cubic Laves phase C15 with lattice constant a = 7.763\,$\AA$ \cite{harris_rare_1965}, a simple cube-on-cube model as (for EuPd$_3$) may not be expected due to the high misfit of $\sim$-7\% with MgO(001).
	No previous study regarding the epitaxial growth of EuPd$_2$ via codeposition exists, so the deposition parameters from the successful EuPd$_3$ growth on MgO(001), as stated above, were used first.
	In contradistinction to the behavior of EuPd$_3$ no change of the out-of-plane growth direction was observed, through changing the MgO substrate annealing temperature.
	Instead, more crystal orientations evolve at lower substrate annealing temperatures around 600\,°C, i.\,e., more intense (100)-, (110)- and (311)-crystal facettes become visible in the high-angle symmetric XRD scans (not shown).
	Furthermore, a lower crystallinity manifests itself in more diffuse RHEED patterns with additional weak dots, indicative of small islands and a very rough surface.
	Best conditions for EuPd$_2$ growth were found to be an annealing temperature of 1000\,°C for one hour, a substrate temperature of 450\,°C during codeposition and an effective slow growth rate around 0.1\,$\AA$/s with stoichiometrically matched Eu and Pd fluxes.
	RHEED azimuthal in-plane scans of a thus prepared thin film exhibit a sixfold symmetry upon rotation with many streaks, including some satellites and also higher order Laue circles (Fig.\,\ref{rheed-eupd2} a).
	Together with the SEM image (Fig.\,\ref{rheed-eupd2} b), revealing triangular formed and weakly coalesced islands, we conclude a pronounced island-like growth behavior of EuPd$_2$ with two in-plane domain variants.
	
	\begin{figure*}[htb]
		\includegraphics[width=\textwidth]{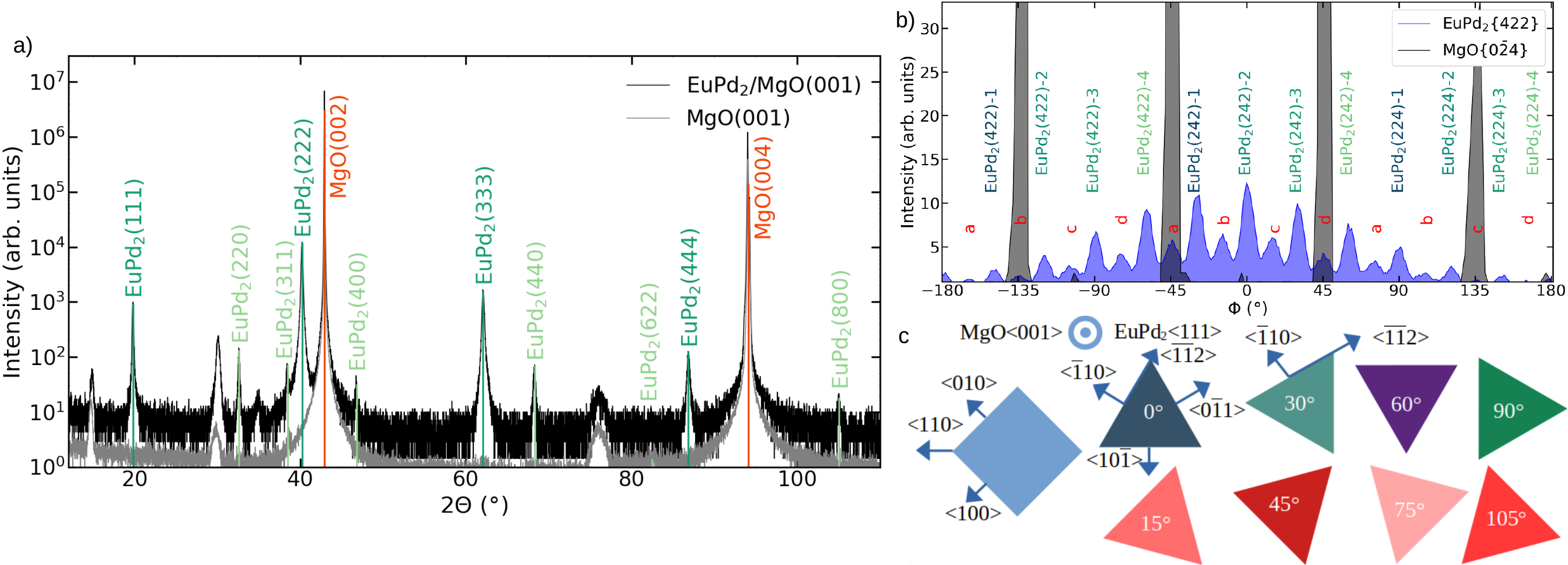}
		\caption{a) Longitudinal symmetric X-ray diffractogram of an epitaxial EuPd$_2$ thin film and a pure MgO(001) substrate. b) $\phi$-scan around sample normal of MgO$\{0\bar{2}4\}$- and EuPd$_2\{422\}$-reflexes in asymmetric geometry. The reflex families for the main in-plane domains are indicated in blueish/greenish colors. Additionally, four minor domains are highlighted by a)-d) in reddish colors. c) Top view sketch of the orientation variants for the main domains (0°, 30°, 60°, 90°) and the minor domains (15°, 45°, 75°, 105°) with respect to the MgO square lattice (light blue) with crystallographic directions as indicated by arrows.}
		\label{xrd-eupd2}
	\end{figure*}
	
	Regarding the symmetric high-angle XRD scan (Fig.\,\ref{xrd-eupd2} a) a strong <111>-out-of-plane orientation for EuPd$_2$ is clearly visible, thus supporting the former findings from the RHEED experiments.
	Although some minor other crystal facets are visible within a logarithmic scale of the intensity, the amount of crystallographic disorder is small.
	Additionally, no other crystalline chemical phase appears, as, e.g., EuPd$_2$Si$_2$ may be expected upon intermixing with the silicon capping layer.
	Besides the EuPd$_2$(111)-reflex small shoulders as Laue oscillations arise, yielding a coherent thickness of L$_c\approx$ 20\,nm.
	Careful inspection of small angle X-ray scans (not shown) do not reveal any Kiessig fringes, pointing to a non-uniform or rough surface, already seen in the SEM images.
	Rocking curve measurements of, e.g., the EuPd$_2$(222) reflex exhibit a full width at half maximum (FWHM) of 0.34°, thus revealing only little mosaic spread and a strong out-of-plane alignment.
	Further $\phi$-scans around the sample's surface normal in asymmetric reflection geometry of the MgO$\{0\bar{2}4\}$- and EuPd$_2\{422\}$-reflexes show regularly spaced peaks 90° and 30° apart, respectively.
	The intensity of all reflexes varies with $\phi$, due to a small sample tilt offset $\sim$0.01$^\circ$ with respect to the diffractometer's $\phi$-circle.
	This offset has no effect on the conclusions regarding the in-plane crystallographic order.
	For EuPd$_2$ different main in-plane orientations, labelled with varying colors and heights in Fig.\,\ref{xrd-eupd2}\,b), are clearly visible and correspond to the four upper orientation models of EuPd$_2$(111) depicted in the sketch in Fig.\,\ref{xrd-eupd2}\,c).
	Furthermore, some small equidistant peaks (labelled with a to d) are visible between those major EuPd$_2$ reflexes, indicating additional in-plane variants, which are sketched below the main domain variants in reddish colors in Fig.\,\ref{xrd-eupd2}\,c). 
	For the main domains, EuPd$_2(10\bar{1})$ aligns parallel with MgO\{110\}, whereas in the case of the minor domains we find EuPd$_2$\{10$\bar{1}$\}$\parallel$MgO\{010\}, leading to an epitaxial relationship for the main in-plane rotational domains: 
	\begin{center}
		MgO\{100\}$\parallel$EuPd$_2$\{111\} \& MgO<110>$\parallel$EuPd$_2$<10$\bar{1}$>
	\end{center}
	And for the minor in-plane domains:
	\begin{center}
		MgO\{100\}$\parallel$EuPd$_2$\{111\} \& MgO<100>$\parallel$EuPd$_2$<10$\bar{1}$>
	\end{center}
	
	\subsubsection{Magnetotransport properties}
	\begin{figure*}[htb]
		\includegraphics[width=\textwidth]{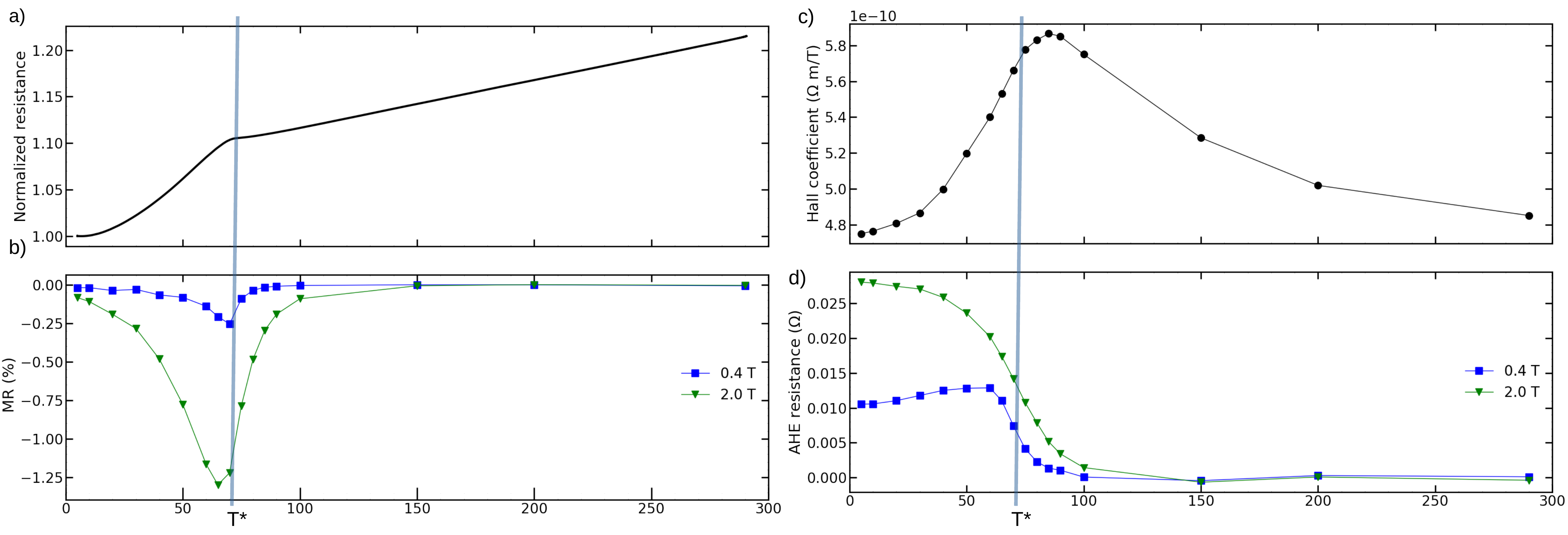}
		\caption{a) Temperature dependence of the electrical 4-wire resistance of the EuPd$_2$ thin film normalized to the value at 5\,K. b) Transverse magnetoresistance MR(B) = $(\rho(T, B) - \rho(T, B = 0))/\rho(T, B = 0)$ with magnetic field oriented out-of-plane. c) Normal Hall coefficient $R_0$ as obtained by a linear fit to the high field data above 2\,T at constant temperatures. d) Anomalous Hall resistance for two different magnetic fields at constant temperatures, as determined after subtracting the field dependent normal Hall contribution. See text for details.}
		\label{transport-EuPd2}
	\end{figure*}
	
	The temperature dependence of the resistivity, shown in Fig.\,\ref{transport-EuPd2} a), reveals a metallic behavior of the lithographically patterned EuPd$_2$ thin film, as expected from single crystal data.
	Above $\sim$90\,K the resistivity scales linearly with temperature, which is indicative of electron-phonon scattering.
	Around 72\,$\pm\,1$\,K a kink is visible, pointing to the onset of magnetic ordering in zero applied field.
	The residual resistivity ratio (RRR) between 280\,K and 3\,K amounts to only $\sim$1.2, which is nearly equal to the values obtained for highly crystalline EuPd$_3$ shown above.
	Using the coherent thickness of 20\,nm, as obtained by analyzing the Laue oscillations, for the calculation of the resistivity for T$\rightarrow 0$ gives a high value around $\rho_0\approx$ 311\,$\mu\Omega$cm.
	Below $\sim$72\,K the resistivity follows $\rho(T) = \rho_0 + A T^2$, with coefficient A =  8.12$\cdot10^{-3}\,\mu\Omega$cm/K$^2$.
	The T$^2$ behavior can be attributed to arising electron-magnon scattering in the magnetically ordered phase \cite{nakamura_magnetic_2016}.
	Due to the weakly connected nature of the islands, the actual in-island resistivity and therefore $\rho_0$ might be much smaller.
	In contrast, EuPd$_2$ single crystals exhibit a higher RRR around 19 with a residual resistivity of $\rho_0\approx1.5\,\mu\Omega$cm for current direction parallel to the a-axis \cite{nakamura_magnetic_2016}.
	In consequence, the electrical transport in epitaxial thin films of EuPd$_2$ is strongly influenced by grain boundary scattering between (weakly) coalesced islands with possible different in-plane orientation.
	Reflecting the distribution of the eight domain variants as obtained by asymmetric XRD scans, a strong increase in resistivity and decrease of RRR is to be expected.\\
	
	\begin{figure*}[htb]
		\includegraphics[width=\textwidth]{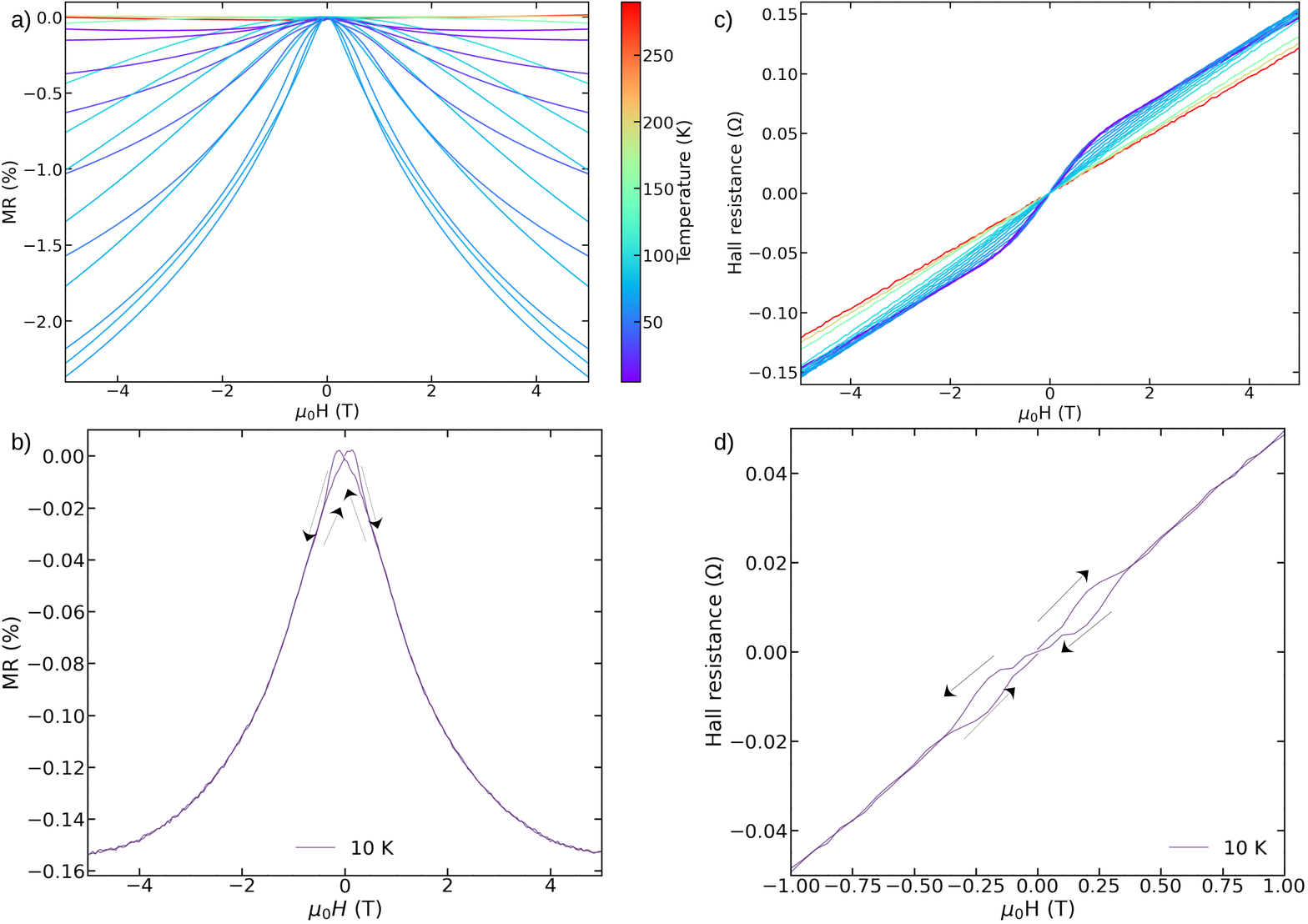}
		\caption{Magnetic field dependence of MR (a) and Hall resistance (c) at constant temperatures, as indicated by the colorbar. MR (b) and Hall resistance (d) at 10\,K as a function of applied magnetic field $\mu_0$H, respectively. Arrows indicate sweep direction of the magnetic field. See text for details.}
		\label{cryo-EuPd2}
	\end{figure*}
	To further investigate the magnetic properties, longitudinal magnetoresistance and Hall effect measurements were performed in an applied field perpendicular to the surface up to 5\,T.
	As is evident from Fig.\,\ref{transport-EuPd2}\,b) a small negative magnetoresistance due to decreasing scattering from magnetic fluctuations or magnetic domain boundaries below about 70\,K is observed.
	At higher temperatures, i.\,e., above 100\,K, no significant magnetoresistance effect is observed.
	Interestingly, the temperature dependence of the magnetoresistance MR(T, B = 0.4\,T), reveals a global minimum at the same magnetic ordering temperature $T^*$ seen in the zero-field resistivity curve (Fig.\,\ref{transport-EuPd2} a).
	Above T$^*$ the MR evolves as downward opened parabola, whose curvature decreases with increasing temperature (Fig.\ref{cryo-EuPd2}\,a).
	In contrast, below T$^*$ the shape of MR exhibits a change in curvature in fields above $\sim$1.5\,T.
	Furthermore, in fields below 0.4\,T hysteresis effects appear upon sweeping the magnetic field, which can be clearly seen in Fig.\,\ref{cryo-EuPd2}\,b).
	
	For ferromagnetic materials the Hall resistance or the transverse resistivity $\rho_{xy}$ may be divided into the ordinary or normal Hall effect (NHE) $\rho^{NHE}_{xy}\propto$ B and the anomalous Hall effect (AHE) $\rho_{xy}^{AHE}\propto$ M with proportionality constants $R_0$ and $R_S$, respectively.
	\[\rho_{xy} = \rho^{NHE}_{xy}(B,T) + \rho_{xy}^{AHE}(M, T) = R_0(T)B + R_S(T)\mu_0M(T)\]
	Due to the saturation behavior of the anomalous Hall effect, corresponding to a saturation of the magnetization, the normal Hall effect contribution can be obtained by a linear fit to the high field data (here above 2\,T, see Fig.\,\ref{cryo-EuPd2} c).
	From the coefficient of the linear slope, we obtain the normal Hall coefficient $R_0(T)$, whose temperature dependence is shown in Fig.\,\ref{transport-EuPd2}\,c).
	Again, a broad maximum near T$^*$ evolves, although the effect of temperature on $R_0$ is small.
	Calculation of the charge carrier density yields around $1\cdot10^{28}$\,m$^{-3}$, whereby the carriers behave hole-like in a single-band mode.
	It should be noted that a hysteretic behavior of $\rho_{xy}$ is observed, below the magnetic ordering temperature and $\mu_0$H\,$\lesssim$\,0.4\,T (Fig.\,\ref{cryo-EuPd2}\,d).
	Subtraction of the normal Hall contribution leaves the anomalous Hall resistance, whose temperature dependence shows the steepest decrease near T$^*$ for different applied fields (Fig.\,\ref{transport-EuPd2}\,d).
	As the magnitude of the anomalous Hall effect increases with decreasing temperature and shows both saturation and hysteresis effects as a function of magnetic field, a ferromagnetic ground state is likely.

	\section{Conclusion}
	In summary, we demonstrated the epitaxial growth of EuPd$_2$ and EuPd$_3$ on Mg(001) substrates using molecular beam epitaxy, with a typical lateral island size of the order of several hundred nm.
	For all thin films investigated, a relaxed growth takes place, such that the cubic a-lattice parameters are equal to their bulk crystal values.
	Depending on the substrate annealing temperature and therefore the surface condition of MgO, two different epitaxial relationships are observed for EuPd$_3$ thin films.
	For annealing temperatures of 600\,°C a simple cube-on-cube relationship is established, showing a defect- or impurity-induced resistance minimum around 15\,K, which is also known from single crystal studies and was attributed to a Kondo effect \cite{takeuchi_contribution_2014}.
	In the high temperature annealing case, a EuPd$_3$(111) orientation is obtained, where the film is nearly fully coherent (about 95\%) in the out-of-plane direction.
	According to RHEED azimuthal scans multiple in-plane domains form, which is also known, e.g., in the simple system Pt/MgO(001) \cite{gatel_morphology_2003}
	In consequence of the higher structural order in mode 2, the residual resistivity is smaller, the RRR is larger and the resistance minimum disappears completely.
	Thin films of EuPd$_3$ thus are less prone to contamination with respect to magnetic impurities as, e.g., potentially Eu$^{2+}$ in EuPd$_3$ single crystals.
	This might be analogous to the case of EuO thin films, whereby reevaporation of unreacted Eu-metal occurs at sufficiently large substrate temperatures on inert YAlO$_3$ \cite{ulbricht_adsorption_2008} or yttria-stabilized zirconia \cite{altendorf_oxygen_2011} substrates, thus avoiding off-stoichiometry or build up of Eu metal clusters under optimum growth conditions.
	An XPS study on the influence of different cleaning methods on the surface conditions on MgO(001) substrates revealed that after vacuum annealing at 700\,°C for 2\,h without any wet-chemical cleaning, still a small amount of Mg(OH)$_2$ remains at the surface \cite{febvrier_wet_2017}.
	The authors claim that further annealing above 900\,°C is necessary to desorb this surface-bound layer, while at the same time calcium segregation to the surface will slowly take place.
	In consequence, the different surface chemisty due to the substrate annealing in the investigated growth modes here, might be responsible for the pronounced change of both, the epitaxial thin film relationship and the amount of crystalline or structural order.
	Even after many years of investigations, the optimum substrate cleaning method for MgO remains under discussion and has to be carefully considered with respect to the growing thin film \cite{febvrier_wet_2017}.
	
	In addition, for the first time EuPd$_2$ thin films could be grown with an epitaxial relationship possessing four main in-plane domains, namely MgO\{100\}$\parallel$EuPd$_2$\{111\} and MgO<110>$\parallel$EuPd$_2$[10$\bar{1}$], as deduced by room temperature RHEED and symmetric and asymmetric XRD experiments.
	Besides these, a minority of the in-plane domains exhibits another crystallographic orientation according to MgO\{100\}$\parallel$EuPd$_2$\{111\} and MgO<100>$\parallel$EuPd$_2$[10$\bar{1}$], which are thus rotated by 15° with respect to the main domains around the growth direction.
	For EuPd$_2$ surface chemistry seems to play a minor role with respect to the epitaxy, although higher annealing temperatures promote a better film crystallinity.
	SEM images indicate an island-like growth mode for EuPd$_2$(111) on MgO(001), which occurs likewise, e.g., for Pd on MgO \cite{renaud_growth_1999} or EuPd$_2$Si$_2$/MgO(001) \cite{koelsch_clamping_2022}, suggesting a high interfacial energy and a strong tendency for dewetting from the MgO substrate.
	Magnetotransport measurements reveal a magnetic phase transition around 72\,K, which corresponds well to the ferromagnetic ordering in single \cite{nakamura_magnetic_2016} and polycrystals \cite{kropp_electric_1979}.
	In combination with the Hall effect investigations, a ferromagnetic ground state of the EuPd$_2$ seems most likely.
	
	To reduce the amount of crystalline in-plane disorder due to rotational domain formation, as seen, e.g., in the resistivity or in RRR, our future studies will focus on the growth of both compounds with the (111) out-of-plane orientation on symmetry-adopted substrates.
	Research along these lines is under way.
	
	\section{Acknowledgment}
	Funded by the Deutsche Forschungsgemeinschaft (DFG, German Research Foundation) - TRR288 - 422213477 (project A04).
	
	\bibliographystyle{unsrt}

\begin{thebibliography}{30}
		
		\bibitem{sampathkumaran_new_1981}
		E.~V. Sampathkumaran, L.~C. Gupta, R.~Vijayaraghavan, K.~V. Gopalakrishnan, R.~G.~Pillay, and H.~G. Devare,
		\newblock A new and unique {Eu}-based mixed valence system: {EuPd}$_{\textrm{2}}${Si}$_{\textrm{2}}$.
		\newblock Journal of Physics C: Solid State Physics (1981), 14(9):L237-L241.
		
		\bibitem{onuki_divalent_2017}
		Y.~Ōnuki, A.~Nakamura, F.~Honda, D.~Aoki, T.~Tekeuchi, M.~Nakashima, Y.~Amako,
		H.~Harima, K.~Matsubayashi, Y.~Uwatoko, S.~Kayama, T.~Kagayama, K.~Shimizu,
		S.~Esakki~Muthu, D.~Braithwaite, B.~Salce, H.~Shiba, T.~Yara, Y.~Ashitomi,
		H.~Akamine, K.~Tomori, M.~Hedo, and T.~Nakama,
		\newblock Divalent, trivalent, and heavy fermion states in {Eu} compounds,
		\newblock Philosophical Magazine (2017), 97(36):3399--3414.
		
		\bibitem{adams_effect_1991}
		D.~M. Adams, A.~E. Heath, H.~Jhans, A.~Norman, and S.~Leonard,
		\newblock The effect of high pressure upon the valence transition in {EuPd}$_{\textrm{2}}${Si}$_{\textrm{2}}$,
		\newblock Journal of Physics: Condensed Matter (1991), 3(29):5465--5468.
		
		\bibitem{mitsuda_field-induced_1997}
		A.~Mitsuda, H.~Wada, M.~Shiga, H.~Aruga~Katori, and T.~Goto,
		\newblock Field-induced valence transition of {Eu}({Pd}$_{\textrm{1-x}}${Pt}$_{\textrm{x}}$)$_{\textrm{2}}${Si}$_{\textrm{2}}$,
		\newblock Physical Review B (1997), 55(18):12474--12479.
		
		\bibitem{kliemt_strong_2022}
		K.~Kliemt, M.~Peters, I.~Reiser, M.~Ocker, F.~Walther, D.-M.~Tran, E.~Cho, M.~Merz, A.~A.~Haghighirad,
		D.~C. Hezel, F. Ritter, and C. Krellner,
		\newblock Influence of the {Pd}-{Si} ratio on the {Valence} {Tansition} in {EuPd}$_{\textrm{2}}${Si}$_{\textrm{2}}$ {Single} {Crystals},
		\newblock Crystal Growth \& Design (2022).
		
		\bibitem{wertheim_final-state_1985}
		G.~K.~Wertheim, E.~V. Sampathkumaran, C.~Laubschat, and G.~Kaindl,
		\newblock Final-state effects in the x-ray photoemission spectrum of {EuPd}$_{\textrm{2}}${Si}$_{\textrm{2}}$,
		\newblock Physical Review B (1985), 31(10):6836--6839.
		
		\bibitem{chatterjee_heavy_2021}
		S.~Chatterjee.
		\newblock Heavy fermion thin films: progress and prospects,
		\newblock Electronic Structure (2021), 3(4):043001.
		
		\bibitem{duriez_structural_1990}
		C.~Duriez, C.~Chapon, C.~R.~Henry, and J.~M.~Rickard,
		\newblock Structural characterization of MgO(100) surfaces,
		\newblock Surface Science (1990), 230(1-3), 123-136.
		
		\bibitem{febvrier_wet_2017}
		Le~Febvrier, J.~Jensen, and P.~Eklund,
		\newblock Wet-cleaning of {MgO}(001): Modification of surface chemistry and effects on thin film growth investigated by x-ray photoelectron spectroscopy and time-of-flight secondary ion mass spectroscopy,
		\newblock Journal of Vacuum Science \& Technology A: Vacuum, Surfaces, and Films (2017), 35(2), 021407.
		
		\bibitem{smith_low-temperature_1968}
		D.~K.~Smith and H.~R.~Leider,
		\newblock Low-temperature thermal expansion of {LiH}, {MgO} and {CaO}.
		\newblock Journal of Applied Crystallography (1968), 1(4):246--249.
		
		\bibitem{gatel_morphology_2003}
		C.~Gatel, P.~Baules, and E.~Snoeck,
		\newblock Morphology of Pt islands grown on MgO (001),
		\newblock Journal of crystal growth (2003), 252(1-3), 424-432.
		
		\bibitem{koelsch_clamping_2022}
		S.~K\"olsch, A.~Schuck, M.~Huth, O.~Fedchenko, D.~Vasilyev, S.~Chernov, O.~Tkach, H.-J.~Elmers, G.~Sch\"onhense, C.~Schl\"uter, T.~R.~F.~Peixoto, A.~Gloskowski, and C.~Krellner,
		\newblock Clamping effect on temperature-induced valence transition in epitaxial {EuPd}$_{\textrm{2}}${Si}$_{\textrm{2}}$ thin films grown on {MgO}(001),
		\newblock Physical Review Materials (2022), 6(11), 115003.
		
		\bibitem{maslankiewicz_europium-palladium_2008}
		P.~Ma{\' s}lankiewicz, Z.~Celinski, and J.~Szade,
		\newblock {Europium}-palladium intermetallic thin layers,
		\newblock Journal of Physics: Condensed Matter (2008), 20(31).
		
		\bibitem{shannon_revised_1976}
		R.~D.~Shannon,
		\newblock Revised effective ionic radii and systematic studies of interatomic distances in halides and chalcogenides,
		\newblock Acta crystallographica section A: crystal physics, diffraction, theoretical and general crystallography (1976), 32(5), 751-767.
		
		\bibitem{doniach_phase_1977}
		S.~Doniach,
		\newblock Phase Diagram for the Kondo Lattice,
		\newblock In: R. D. Parks (eds) Valence Instabilities and Related Narrow-Band Phenomena (1977), Springer, Boston.
		
		\bibitem{villars_pearsons_1991}
		P.~Villars and L.~D.~Calvert,
		\newblock Pearson’s Handbook of Crystallographic Data for Intermetallic Phases (1991), 2nd edition, vol. 4 
		\newblock Materials Park, OH: ASM International
		
		\bibitem{matula_electrical_1979}
		R.~A.~Matula,
		\newblock Electrical resistivity of copper, gold, palladium, and silver,
		\newblock Journal of Physical and Chemical Reference Data (1979), 8(4), 1147-1298.
		
		\bibitem{nakamura_magnetic_2016}
		A.~Nakamura, H.~Akamine, Y.~Ashitomi, F.~Honda, D.~Aoki, T.~Takeuchi, 
		K.~Matsubayashi, Y.~Uwatoko, Y.~Tatetsu, T.~Maehira, M.~Hedo, T.~Nakama, Y.~\& $\bar{O}$nuki,
		\newblock Magnetic and {Fermi} {Surface} {Properties} of {Ferromagnets} {EuPd}$_{\textrm{2}}$ and {EuPt}$_{\textrm{2}}$,
		\newblock Journal of the Physical Society of Japan (2016), 85(8), 084705.
		
		\bibitem{takeuchi_contribution_2014}
		T.~Takeuchi, A.~Nakamura, M.~Hedo, T.~Nakama, and Y.~\& $\bar{O}$nuki,
		\newblock {Contribution} of {J}-multiplet levels to the physical properties of {EuPd}$_{\textrm{3}}$ with the trivalent electronic state,
		\newblock Journal of the Physical Society of Japan (2014), 83(11), 114001.
		
		\bibitem{gardner_the_1972}
		W.~E.~Gardner, J.~Penfold, T.~F.~Smith, and I.~R.~Harris,
		\newblock The magnetic properties of rare earth-{Pd}$_{\textrm{3}}$ phases,
		\newblock Journal of Physics F: Metal Physics (1971), 2(1), 133.
		
		\bibitem{harris_rare_1965}
		I.~R.~Harris, and G.~V.~Raynor,
		\newblock Rare-earth intermediate phases: II. {Phases} formed with palladium,
		\newblock Journal of the Less Common Metals (1965), 9(4), 263-269.
		
		\bibitem{harris_study_1971}
		I.~R.~Harris, G.~Longworth,
		\newblock A study of the lattice spacings, magnetic susceptibilities and 151{Eu} {Mössbauer} spectra of some palladium-europium alloys,
		\newblock Journal of the Less Common Metals (1971), 23(3), 281-292.
		
		\bibitem{iandelli_europium_1974}
		A.~Iandelli, A.~Palenzona,
		\newblock The europium-palladium system,
		\newblock Journal of the Less Common Metals (1974), 38(1), 1-7.
		
		\bibitem{bertran_growth_1991}
		F.~Bertran, T.~Gourieux, G.~Krill, M.~Alnot, J.~J.~Ehrhardt, and W.~Felsch,
		\newblock Growth of {Eu} on {Pd}(111): {AES}, photoemission and {RHEED} studies,
		\newblock Surface Science Letters (1991), 245(1-2), L163-L169.
		
		\bibitem{bertran_growth_1992}
		F.~Bertran, T.~Gourieux, G.~Krill, M.~F.~Ravet-Krill, M.~Alnot, J.~J.~Ehrhardt, and W.~Felsch,
		\newblock Growth of {Eu} on {Pd}(111) studied by x-ray and uv photoemission and crystallographic properties as determined by reflection-high-energy-electron-diffraction and x-ray-diffraction studies,
		\newblock Physical Review B (1992), 46(12), 7829.
		
		\bibitem{bertran_the_1992}
		F.~Bertran, T.~Gourieux, G.~Krill, M.~F.~Ravet-Krill, M.~Alnot, J.~J.~Ehrhardt, and W.~Felsch,
		\newblock The {Eu/Pd}(111) interface: spectroscopic and structural studies,
		\newblock Surface science (1992), 269, 731-736.
		
		\bibitem{wickman_moessbauer_1967}
		H.~H.~Wickman, J.~H.~Wernick, R.~C.~Sherwood, and C.~F.~Wagner,
		\newblock M\"ossbauer and magnetic properties of several europium intermetallic compounds,
		\newblock Bell Telephone Labs. (1967), Inc., Murray Hill, NJ.
		
		\bibitem{kropp_electric_1979}
		H.~Kropp, E.~Dormann, and K.~H.~J.~Buschow,
		\newblock Electric field gradient in cubic intermetallic europium compounds with unstable europium valence,
		\newblock Solid State Communications (1979), 32(7), 507-510.
		
		\bibitem{kropp_indirect_1979}
		H.~Kropp, W.~Zipf, E.~Dormann, and K.~H.~J.~Buschow,
		\newblock Indirect exchange in intermetallic europium compounds,
		\newblock Journal of Magnetism and Magnetic Materials (1979), 13(1-2), 224-230.
		
		\bibitem{ulbricht_adsorption_2008}
		R.~W.~Ulbricht, A.~Schmehl, T.~Heeg, J.~Schubert, and D.~G.~Schlom,
		\newblock Adsorption-controlled growth of {EuO} by molecular-beam epitaxy,
		\newblock Applied physics letters (2008), 93(10), 102105.
		
		\bibitem{altendorf_oxygen_2011}
		S.~G.~Altendorf, A.~Efimenko, V.~Oliana, H.~Kierspel, A.~D.~Rata, and L.~H.~Tjeng,
		\newblock Oxygen off-stoichiometry and phase separation in {EuO} thin films,
		\newblock Physical Review B (2011), 84(15), 155442.
		
		\bibitem{mimura_temperature-induced_2011}
		K.~Mimura, T.~Uozumi, T.~Ishizu, S.~Motonami, H.~Sato,
		Y.~Utsumi, S.~Ueda, A.~Mitsuda, K.~Shimada, Y.~Taguchi, 
		Y.~Yamashita, H.~Yoshikawa, H.~Namatame, 
		M.~Taniguchi, and K.~Kobayashi,
		\newblock Temperature-{Induced} {Valence} {Transition} of
		{EuPd}$_{\textrm{2}}${Si}$_{\textrm{2}}$ {Studied} by {Hard} {X}-ray {Photoelectron} {Spectroscopy},
		\newblock Japanese Journal of Applied Physics (2011), 50(5):05FD03.
		
		\bibitem{mimura_temperature_2004}
		K.~Mimura, Y.~Taguchi, S.~Fukuda, A.~Mitsuda, J.~Sakurai, K.~Ichikawa, and O.~Aita,
		\newblock Temperature dependence of {Eu} 4f states in {EuPd}$_{\textrm{2}}${Si}$_{\textrm{2}}$: bulk-sensitive high-resolution photoemission study,
		\newblock Physica B: Condensed Matter (2004), 351(3-4):292--294.
		
		\bibitem{wada_first-order_1996}
		H.~Wada, A.~Mitsuda, M.~Shiga, H.~A.~Katori, and T.~ Goto,
		\newblock First-order valence transition of {EuPd}$_{\textrm{2}}${Si}$_{\textrm{2}}$ induced by high magnetic fields,
		\newblock Journal of the Physical Society of Japan (1996), 65(11), 3471-3473.
		
		\bibitem{renaud_growth_1999}
		G.~Renaud, A.~Barbier, and O.~Robach,
		\newblock Growth, structure, and morphology of the {Pd}/{MgO}(001) interface: {Epitaxial} site and interfacial distance,
		\newblock Physical Review B (1999), 60(8):5872--5882.
	\end{thebibliography}

\end{document}